\documentclass[conference]{IEEEtran}
\IEEEoverridecommandlockouts
\pdfoutput=1
\usepackage{rotating,setspace,latexsym,amsmath,amssymb,bm}
\usepackage{cite}
\usepackage{algpseudocode}
\usepackage{algorithm}
\usepackage{algorithmicx}
\usepackage{xfrac}
\usepackage{amsthm}
\usepackage{subfig}
\usepackage{graphicx}
\usepackage{booktabs} 
\usepackage{caption}
\usepackage{hyperref}

\begin{document}
\pagestyle{empty}
\title{On the Tradeoff between Energy Harvesting and Caching in Wireless Networks}
\author{\IEEEauthorblockN{Akshay Kumar and Walid Saad}
\IEEEauthorblockA{Wireless@VT, Bradley Department of Electrical and Computer Engineering,\\
Virginia Tech., Blacksburg, Virginia (USA)\\
Email:\{akshay2,walids\}@vt.edu}
\thanks{This research was supported by the U.S. National Science Foundation under Grant CNS-1460333.}
}
\maketitle 
\begin{abstract}
Self-powered, energy harvesting small cell base stations (SBS) are expected to be an integral part of next-generation wireless networks. However, due to uncertainties in harvested energy, it is necessary to adopt energy efficient power control schemes to reduce an SBSs' energy consumption and thus ensure quality-of-service (QoS) for users. Such energy-efficient design can also be done via the use of content caching which reduces the usage of the capacity-limited SBS backhaul. of popular content at SBS can also prove beneficial in this regard by reducing the backhaul usage. In this paper, an online energy efficient power control scheme is developed for an energy harvesting SBS equipped with a wireless backhaul and local storage. In our model, energy arrivals are assumed to be Poisson distributed and the popularity distribution of requested content is modeled using Zipf's law. The power control problem is formulated as a (discounted) infinite horizon dynamic programming problem and solved numerically using the value iteration algorithm. Using simulations, we provide valuable insights on the impact of energy harvesting and caching on the energy and sum-throughput performance of the SBS as the network size is varied. Our results also show that the size of cache and energy harvesting equipment at the SBS can be traded off, while still meeting the desired system performance.
\end{abstract}

\section{Introduction}
The exponential increase in mobile data traffic has led to the advent of small cell networks as a promising solution to meet the ever increasing demand for the wireless spectrum \cite{NSN11, Damn11}. From the perspective of mobile operators, rural areas present a great business opportunity for deploying small cells due to increased coverage, subscriber base, and paid usage per subscriber \cite{smallCellForum}. However, there are two main challenges in rural small cell deployment: energy efficiency and backhaul availability. Many locations, especially in emerging markets, have limited or no access to the conventional power grid and, thus they need to harvest energy from other, renewable sources, such as solar. The amount of energy harvested fluctuates over time resulting in potential power outages at the SBS and thus degrading the QoS for the users. Therefore, it is imperative to design and implement energy-efficient transmission schemes.  Moreover, rural small cells are often outside the normal economic reach of existing operator backhaul infrastructure. This significantly increases the backhaul cost and thus limits the backhaul capacity, making it difficult to deliver high speed mobile broadband coverage. 

Resource allocation for energy harvesting wireless communication systems has been studied extensively \cite{Matthias13, Sabella14, zhangLau14, huang14, KwanLo13, gong14}. While some of the resource allocation schemes \cite{Matthias13, Sabella14} focus on maximizing energy efficiency for a small cell network, they do not consider energy harvesting. Other works focus on optimizing either sum throughput \cite{huang14} or delay performance \cite{zhangLau14} rather than the energy efficiency of renewable energy powered systems. Only recently there has been some work \cite{KwanLo13, gong14} done on minimizing power consumption in renewable energy powered cellular networks. The authors in \cite{gong14} formulated an optimization problem to adapt the BSs' on-off states, active resource (spectrum) blocks and renewable energy allocation to minimize the average grid power consumption while satisfying users' QoS requirements. However, the limitation of this work is that it assumes that the energy arrivals are known non-causally and BSs' use \emph{on}-\emph{off} power control rather than using a variable power control. Also, both \cite{KwanLo13} and \cite{gong14} assume hybrid BSs' with access to both power grid and renewable energy. However, this may not be a realistic assumption for rural small cells, which have little to no-access to power grid. Therefore, it is desirable to develop \emph{online} energy-efficient power control schemes for purely renewable energy powered small cell networks, while ensuring a best effort QoS for users.  

Beyond adopting energy-efficient transmission schemes, one can also cache popular content at the level of the base stations to further reduce the amount of energy consumed in the network. Caching of popular content reduces the backhaul usage for downloading user data and thus ensures QoS during peak traffic demands\footnote{Caching of popular user data is particularly beneficial for SBS because they typically have weak backhaul links with limited bandwidth.}. The idea of edge caching was first proposed in \cite{molisch13, Golrezaei12} in which a new architecture for wireless video distribution based on distributed caching of popular content at the level of small cells. It studied the problem of optimal content placement across distributed caches such that the expected sum delay in downloading users data is minimized. This work was extended in \cite{Tadrous13, Caire13} to include the notion of proactive resource allocation exploiting the predictability of user behavior for load balancing. Recently, the authors in \cite{pingyod14} proposed a low complexity scheme for updating cached content in a small cell network. Nevertheless, while interesting, these works primarily focus only on content placement and update so as to minimize the total delay in downloading the user data. The impact of caching on the energy consumption and backhaul usage for the small cell network is yet to be explored. This is particularly important for renewable energy powered small cell networks with limited battery capacity and backhaul bandwidth.
\begin{figure*}[!t]
\centering
\includegraphics[scale=0.45]{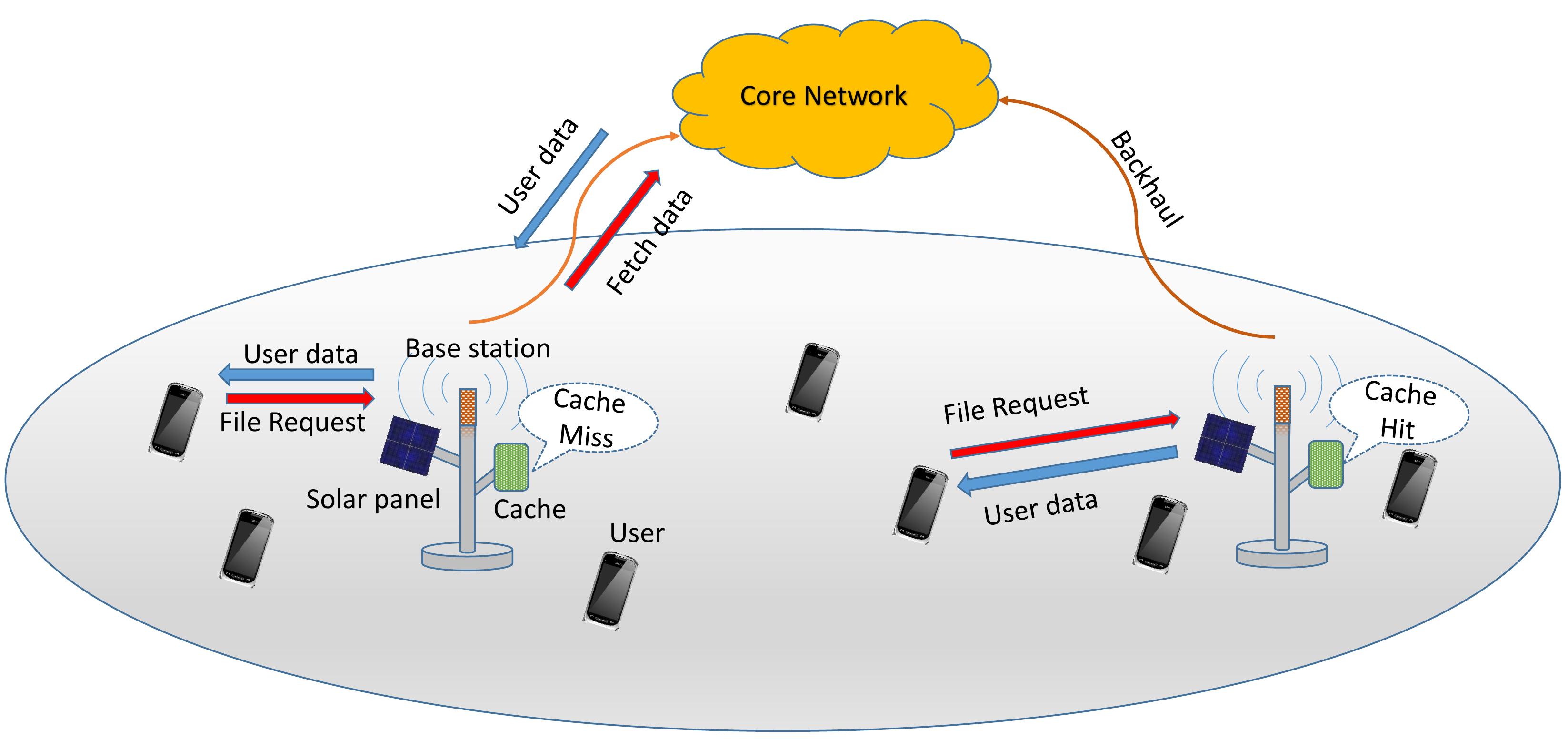}%
\caption{System Model}%
\label{sysModelFig}%
\end{figure*}

The main contribution of this paper is to develop an online energy-efficient power control scheme for an energy harvesting SBS that is endowed with caching capabilities and a wireless, capacity-limited backhaul. Energy arrival are modeled as Poisson process and the popularity distribution of user data is modeled using Zipf's law. The utility function of the SBS is a generalized energy-efficiency function that alters the degree of energy efficiency depending upon the current battery level. We formulate the power control problem at the SBS as a (discounted) infinite horizon dynamic programming (DP) problem that maximizes the average energy efficiency of system subject to certain power constraints for the SBS. The DP problem is solved numerically using a proposed value iteration algorithm. Simulation results indicate that increasing cache size and/or energy harvesting capability increases energy availability at SBS. Interestingly, this comes at a gain in sum throughput for the proposed scheme as compared to the sum throughput maximization scheme. Finally, we show that for a given performance criteria, it is possible to trade off the energy harvesting capability for cache size.

The rest of the paper is organized as follows. Section~\ref{sysModel} introduces the system model. In Section~\ref{probForm}, we formulate the optimization problem to determine the transmit powers on downlink and backhaul. We then propose a value iteration algorithm to solve this problem. Section~\ref{Results} presents simulation results. Finally Section~\ref{concl} draws some conclusions.

\section{System Model}
\label{sysModel}
Figure~\ref{sysModelFig} shows the system model of an SBS serving $N_u$ mobile users. The SBS is equipped with a local cache of size $M$ and is connected to the core network using a wireless (licensed spectrum) backhaul link. We also assume that the SBS is self-powered and it obtains its power supply through an energy harvesting device that is connected to a rechargeable battery with capacity $E_{\textnormal{max}}$. 

In this model our goal is to study the impact of caching and energy harvesting on resource allocation in a SBS. In our case, resources consist of power and bandwidth allocated to the users. The scheduler at the SBS performs resource allocation at the start of each time slot of duration $T$ seconds. Let $E_k \in [0, E_{\text{max}}]$ be the energy remaining in the battery at the start of the $k^{\textnormal{th}}$ time slot. Then the evolution of $E_k$ over time can be written recursively as,
\begin{equation}
\label{energyUpdate}
E_{k+1} = \text{min}(E_k-P_k T+\tilde{E}_k,E_{\text{max}}),~~\forall~k = 0, 1, 2, ...
\end{equation}
Here, $\tilde{E}_k$ is a random variable denoting the energy harvested during the $k^\textnormal{th}$ slot. The energy arrivals are typically modeled as a Poisson process with mean $\lambda$ \cite{Ozel11, Dhillon14, Gelenbe14}. Let $q$ be the amount of energy harvested at each arrival, which depends on the capability of the energy harvesting device present at the SBS. So $\tilde{E}_k = N_a q$, where $N_a$ is the number of arrivals in time $T$ with a mean value of $\lambda T$. The variable $P_k \in [0, \text{min}(P_{\textnormal{max}}, \frac{E_k}{T})]$ in \eqref{energyUpdate} represents the total transmit power of the SBS during the $k^{\textnormal{th}}$ slot. $P_{\textnormal{max}}$ is the maximum transmit power constraint on the SBS, but the actual upper bound is $\text{min}(P_{\textnormal{max}}, \frac{E_k}{T})$ because the energy remaining in the battery cannot be negative at any instant. The total power $P_k$ can be written in terms of its constituents as,
\begin{equation}
P_k=\sum_{i=1}^{N_u}P_{i,k} + \epsilon P_{b,k},
\label{power}
\end{equation}
where $P_{i,k}$ is the transmit power on downlink for user $i$ during the $k^{\textnormal{th}}$ slot. Similarly, $P_{b,k}$ is the SBS transmit power spent on fetching the user requested content from the core network using the backhaul. The backhaul link is accessed with a probability $\epsilon$, which is related\footnote{The backhaul is accessed for fetching the files from the core network if atleast one of files requested by the $N_u$ users is not present in the cache. The probability that all the users find their requested files in cache is ${(1-\epsilon_m)}^{N_u}$. Therefore probability of using backhaul is $1-{(1-\epsilon_m)}^{N_u}$.} to the probability of cache miss, $\epsilon_m$, such that $\epsilon=1-{(1-\epsilon_m)}^{N_u}$. We assume that the SBS caches the $M$ most popular files among the users, because the probability of users requesting these files is more than other files and thus minimizes the backhaul usage for fetching the requested content from the core. Therefore, caching the popular data is likely to result in higher energy savings and lower latency than just randomly caching user content. We model the content popularity distribution using Zipf's law which states that the frequency of an element (file) of rank $j$ in a set of $R$ elements is given as,
\begin{equation}
f(j;s,R)=\frac{1/j^s}{\sum_{r=1}^{R} 1/n^s},
\label{Zipf}
\end{equation}
where $s>1$ is a decay constant characterizing the \emph{peakiness} of the distribution: a large value indicating that a very small number of files account for majority of data traffic. The rank of a file indicates its relative popularity, with a lower rank indicating higher popularity. The probability of cache miss, $\epsilon$, is the probability that one of the files that has not been cached is requested by the user and can be calculated as,
\begin{equation}
\epsilon =\sum_{j=M+1}^R f(j;s,R) = \frac{\sum_{j=M+1}^R 1/j^s}{\sum_{r=1}^{R} 1/n^s}.
\label{probMiss}
\end{equation}

For a given resource (bandwidth, power) allocation and battery level, we define the utility of the SBS at time slot $k$ as,
\begin{equation}
U_k (\boldsymbol{w}_k, \boldsymbol{p}_k, \boldsymbol{h}_k, E_k) = \frac{\sum_{i=1}^{N_u} W_{i,k} \text{log}_2 (1+\gamma_{i,k})}{{(\sum_{i=1}^{N_u} P_{i,k})}^{g(E_k)}},
\label{utility}
\end{equation}
where $W_{i,k}$ is the bandwidth allocated to the $i^{\textnormal{th}}$ user and the backhaul respectively during the $k^{\textnormal{th}}$ slot\footnote{In LTE, the scheduler allocates the bandwidth to a UE at each TTI in multiples of Physical Resource Block (PRB), where a PRB comprises of 12 subcarriers each 15~kHz wide.}. For simplicity, we have $\boldsymbol{w}_k = [W_{1,k}, W_{2,k},\cdots, W_{N_u,k}]$ and $\boldsymbol{p}_k = [P_{1,k}, P_{2,k},\cdots, P_{N_u,k}]$. The variables $\gamma_{i,k}$ in \eqref{utility} denotes the received signal-to-noise ratio (SNR) at the $i^{\textnormal{th}}$ user at $k^{\textnormal{th}}$ slot, and are given by,
\begin{equation}
\gamma_{i,k} = \frac{P_{i,k}{\|h_{i,k}\|}^2}{d_i^{\alpha} {\sigma}^2},
\label{gamma}
\end{equation}
where $h_{i,k}$ denotes the downlink channel between the SBS and $i^{\textnormal{th}}$ user. $d_i$ is the distance\footnote{We assume that the users are stationary over the time duration of interest.} between the SBS and the $i^{\text{th}}$ user and $\alpha$ is the path loss exponent. All the channels are assumed to be i.i.d., Rayleigh fading channels. ${\sigma}^2$ is the variance of the Gaussian noise.

When the user-requested data is not present in the local cache at SBS, then the requests must be forwarded to the core network via backhaul. When it receives the data from core network, the SBS transmits it back on the downlink\footnote{The contents of the cache may be updated when new data is downloaded using backhaul, if file popularity distribution has changed over the course of time.}. Let $\gamma_{b,k}$ denote the received SNR at the core network when the backhaul is used to fetch the files from the core network. $\gamma_{b,k}$ can be written as,
\begin{equation}
\gamma_{b,k} = \frac{P_{b,k}{\|h_{b,k}\|}^2}{d_b^{\alpha} {\sigma}^2},
\label{gamma_backhaul}
\end{equation}
where $h_{b,k}$ denotes the Rayleigh fading channel between the SBS and the core network. $d_b$ is the distance between the SBS and the core network. The SBS transmit power on backhaul, $P_{b,k}$, should be large enough so that $\gamma_{b,k} \geq \gamma_{\text{min}}$ at each time slot $k$. Otherwise, the receiver at core network cannot correctly decode the files requested by the users.

The function $g(E_k)$ in \eqref{utility} alters the degree of energy efficiency in next time slot depending upon the current battery level, $E_k$. Although many functions can be used, we define $g(E_k)$ as,
\begin{equation}
g(E_k) = a + \frac{(b-a)E_k}{E_{max}},
\label{funcg}
\end{equation}
where $a$ and $b$ are constants $(a>b>0)$, that determine the extent to which $E_k$ influences energy efficiency. We choose this function because it is desirable to be more energy efficient when battery level is low, due to uncertainty in harvested energy. Clearly, the function $g(E_k)$ in \eqref{funcg} satisfies this and is a non-increasing function of $E_k$. Also, both $g(E_k)$ and $U_k$ are well-defined at all values of $E_k$. 

We now determine the average utility of the SBS, $U_k^{\text{avg}} (\boldsymbol{w}_k, \boldsymbol{p}_k, E_k)$, by averaging over all possible channel realizations as following,
\begin{align}
U_k^{\text{avg}} (\boldsymbol{w}_k, \boldsymbol{p}_k, E_k) &= \mathbb{E}_{\boldsymbol{h}_k} \left(U_k (\boldsymbol{w}_k, \boldsymbol{p}_k, \boldsymbol{h}_k, E_k)\right), \label{avgUtil}\\
&= \frac{\mathbb{E}_{\boldsymbol{h}_k} \left(\sum_{i=1}^{N_u} W_{i,k}\text{log}_2 (1+\frac{P_{i,k}{\|h_{i,k}\|}^2}{d_i^{\alpha} {\sigma}^2})\right)}{{[\sum_{i=1}^{N_u} P_{i,k}]}^{g(E_k)}},\label{avgUtilLater}
\end{align}
where $\mathbb{E}_{\boldsymbol{h}_k}(.)$ denotes the expectation operation over the channel vector $\boldsymbol{h}_k = [h_{1,k}, h_{2,k}, \cdots, h_{N_u,k}]$. Equation~\eqref{avgUtilLater} follows from substituting \eqref{utility} and \eqref{gamma} in \eqref{avgUtil}. Since we assumed the channel to be Rayleigh fading, the squared norm of the channel, $H_{i,k} = {\|h_{i,k}\|}^2$, is exponentially distributed with some rate parameter, $\mu~\forall~i, k$, and is given by,
\begin{equation}
f_{H_{i,k}}(x) = \mu \text{e}^{-\mu x}~\forall~i, k.
\label{expDist}
\end{equation}
Then using \eqref{avgUtilLater} and \eqref{expDist}, we have,
\begin{align}
U_k^{\text{avg}} (\boldsymbol{w}_k, \boldsymbol{p}_k, E_k) &= \frac{\sum_{i=1}^{N_u} W_{i,k}\int_0^{\infty} \text{log}_2 (1+\frac{P_{i,k}x}{d_i^{\alpha} {\sigma}^2}) \text{e}^{-\mu x} dx}{\frac{1}{\mu}{[\sum_{i=1}^{N_u} P_{i,k}]}^{g(E_k)}}.
\label{avgUtil2}
\end{align}
Let us solve the integral $I = \int_0^{\infty} \text{log}_2 (1+\frac{P_{i,k}x}{d_i^{\alpha} {\sigma}^2}) \text{e}^{-\mu x} dx$. Let $r=\frac{P_{i,k}}{d_i^{\alpha} {\sigma}^2}$. Applying integration by parts to $I$, we have,
\begin{align}
I &= \frac{1}{\mu~\text{ln}(2)}\left(\left.-\text{ln}(1+rx)\text{e}^{-\mu x}\right|^{\infty}_{0} + \int_0^{\infty} \frac{r}{1+rx} \text{e}^{-\mu x} dx\right),\label{eq1}\\
 &= \frac{\text{e}^{\frac{\mu}{r}}}{\mu~\text{ln}(2)}\int_{\frac{\mu}{r}}^{\infty} \frac{\text{e}^{-z}}{z} dz = \frac{\text{e}^{\frac{\mu d_i^{\alpha} {\sigma}^2}{P_{i,k}}} \Gamma(0,\frac{\mu d_i^{\alpha}{\sigma}^2}{P_{i,k}})}{\mu~\text{ln}(2)} \label{eq2},
\end{align} 
where \eqref{eq2} follows from \eqref{eq1} by noting that $\lim_{x\rightarrow \infty} -\text{ln}(1+rx)\text{e}^{-\mu x} = 0$ (using L'Hospital's rule) and by setting $z=\mu (x+\frac{1}{r})$. Then noting that the integral is a upper incomplete Gamma function, $\Gamma(0,\frac{\mu}{r})$ and substituting back the value of $r$, we get the final expression in \eqref{eq2}. Substituting the solution of the integral in \eqref{avgUtil2} we get,
\begin{equation}
U_k^{\text{avg}} (\boldsymbol{w}_k, \boldsymbol{p}_k, E_k) = \frac{\mu \sum_{i=1}^{N_u} \left(W_{i,k} \text{e}^{\frac{\mu d_i^{\alpha} {\sigma}^2}{P_{i,k}}} \Gamma(0,\frac{\mu d_i^{\alpha} {\sigma}^2}{P_{i,k}}) \right)}{\text{ln}(2){[\sum_{i=1}^{N_u} P_{i,k}]}^{g(E_k)}}.
\label{finalAvgUtil}
\end{equation}

\section{Problem Formulation and Solution Methodology}
\label{probForm}
In this section, we first formulate the optimal power control problem at the SBS as an infinite horizon dynamic programming\footnote{The SBS continually serves the users that are associated to it as long as the energy in the battery is above a particular threshold, so it is reasonable to consider an infinite-horizon problem.}(DP) problem and then present a solution approach for finding the optimal transmit power at each time slot.

For the system model described in the last section, the problem at the SBS is that of optimal power control $\boldsymbol{p}_k^{*}$ for given energy $E_k$ and channel state $\boldsymbol{h}_k$ at time slot $k$. It is determined by solving for the power control scheme that maximizes the sum of current and (discounted) future utility. Mathematically this is given as,
\begin{align}
\boldsymbol{p}_k^{*}(E_k, \boldsymbol{h}_k) &= \operatorname*{arg\,max}_{\boldsymbol{p}_k}~~U_k (\boldsymbol{w}_k, \boldsymbol{p}_k, E_k, \boldsymbol{h}_k) \nonumber \\ &+ \delta \sum_{E_{k+1}} \mathcal{P}(E_{k+1}/E_k,\boldsymbol{p}_k) V_{\pi^{\dagger}}(E_{k+1}), \label{optimalPowerFinal}
\\ \text{such that} \nonumber \\
& p_{i, k} \geq 0, \forall~i=1, 2, \cdots, N_u, , \label{c1}\\
& \sum_{i=1}^{N_u}P_{i,k} + \epsilon P_{b,k} \leq \text{min}(P_{\text{max}}, \frac{E_k}{T}). \label{c2}	
\end{align}
The backhaul power, $P_{b,k}$ in \eqref{c2} is set such that $\gamma_{b,k} = \gamma_{\text{min}}$. If it turns out that $\frac{\text{min}(P_{\text{max}}, \frac{E_k}{T}) {\|h_{b,k}\|}^2}{d_b^{\alpha} {\sigma}^2} < \gamma_{\text{min}}$, then set $P_{b,k} = P_{i,k} = 0$.

$\delta \in [0,1)$ is a discount factor for the utility of future states. $\mathcal{P}(E_{k+1}/E_k,\boldsymbol{p}_k)$ denotes the transition probability from energy state $E_k$ to $E_{k+1}$ when the transmit power in the $k^{\text{th}}$ slot is $\boldsymbol{p}_k$. $V_{\pi^{\dagger}}(E_{k+1})$ is the \emph{value} of state $E_{k+1}$ under the transmission policy $\pi^{\dagger}(E_k)$ and is defined iteratively as,
\begin{align}
V_{\pi^{\dagger}}(E_k) &= U_k^{\text{avg}} (\boldsymbol{w}_k, \boldsymbol{p}_k, E_k) \nonumber \\ &+ \delta \sum_{E_{k+1}} \mathcal{P}(E_{k+1}/E_k,\boldsymbol{p}_k) V_{\pi^{\dagger}}(E_{k+1}), \label{obj} 	
\end{align}
where $\boldsymbol{p}_k=\pi^{\dagger}(E_k)$. Note that the value function uses the average utility function defined in \eqref{avgUtil}.

The policy $\pi^{\dagger}(E_k)$ is the one that maximizes the value function $V_{\pi}(E_k)$, subject to certain power constraints. So we have,
\begin{align}
\pi^{\dagger}(E_k) &= \operatorname*{arg\,max}_{\boldsymbol{p}_k}~~U_k^{\text{avg}} (\boldsymbol{w}_k, \boldsymbol{p}_k, E_k) \nonumber \\ &+ \delta \sum_{E_{k+1}} \mathcal{P}(E_{k+1}/E_k,\boldsymbol{p}_k) V_{\pi^{\dagger}}(E_{k+1}), \label{optimalPolicy}
\\ \text{subject to}~~\eqref{c1}, \eqref{c2}. \nonumber
\end{align}
The backhaul power, $P_{b,k}$ in \eqref{c2} is set such that $\frac{P_{b,k}}{\mu d_b^{\alpha} {\sigma}^2} = \gamma_{\text{min}}$.

It is not possible to obtain a closed form solution to \eqref{optimalPowerFinal}, because the objective function depends on the \emph{value} of next state which in turn depends on the \emph{value} of next-to-next stage and so on. Therefore, we need to solve the DP problem numerically. Policy iteration and value iteration are the two popular algorithms for numerically solving the DP. Policy iteration converges quickly, but it requires search over set of all feasible policies which is typically a very large set. On the other hand, value iteration iteratively updates the value of each state by determining the actions that would maximize the value of that state. Therefore, the search is over a much smaller set. The drawback is slow convergence. We decided to use value iteration to solve the DP in \eqref{optimalPowerFinal}. However, before doing that, we would need to discretize the state space and action space. The detailed algorithm is described in Algorithm~1.
\begin{algorithm}[!t]
\caption{Proposed Value Iteration Algorithm}
 \textbf{Procedure:} \text{Value\_Iteration} $(\mathcal{E}, A,\mathcal{P}, U^{\text{avg}}, \theta)$\\
 \textbf{Inputs} \\
	\text{$\mathcal{E}=\{0, 1, \cdots, E_{\text{max}}\}$ is the set of all energy states}\\
	\text{$A$ is the set of all transmit power vectors, $\boldsymbol{p}$}\\
	\text{$\mathcal{P}$ is the state transition function specifying $\mathcal{P}(E^{'}/E,\boldsymbol{p})$}\\
	\text{$U^{\text{avg}}$ is the utility function $U(\boldsymbol{p}, E, \boldsymbol{h})$}\\
	\text{$\theta$ is a threshold, $\theta>0$}\\
	 \textbf{Outputs} \\
	\text{$\pi^{\dagger}[\mathcal{E}]$ is optimal policy}\\
	\text{$V[\mathcal{E}]$ is the value function}\\
	 \textbf{Local} \\
	\text{real array $V_m[\mathcal{E}]$ is a sequence of value functions}\\
	\text{action array $\pi^{\dagger}[\mathcal{E}]$}\\
	\textbf{Begin Algorithm:}
		\begin{algorithmic} 
		\label{valueIterAlgo}
		\State $V_0[E] \leftarrow 0~~\forall E\in\mathcal{E}$
		\State $m\leftarrow0$
		\Repeat
		\State $m\leftarrow m+1$
		\For{$E=0$ to $E_{\text{max}}$} \Comment{Update value of each state}
		\State $V_{m}(E) = \text{max}_{\boldsymbol{p}\in A}~~U^{\text{avg}}(\boldsymbol{w}_k, \boldsymbol{p}, E) + \delta \sum_{E^{'}} \mathcal{P}(E^{'}/E,\boldsymbol{p}) V_{m-1}(E{'})$
		\EndFor 
		\Until{$\forall E,~ |V_m[E]-V_{m-1}[E]|<\theta$}
				\For{$E=0$ to $E_{\text{max}}$} \Comment{Determine the optimal policy}
		\State $\pi^{\dagger}(E) = \operatorname*{arg\,max}_{\boldsymbol{p}_k\in A} U^{\text{avg}}(\boldsymbol{w}_k, \boldsymbol{p}, E) + \delta \sum_{E^{'}} \mathcal{P}(E^{'}/E,\boldsymbol{p}) V_{m}(E{'})$
		\EndFor \\
		\Return $\pi^{\dagger},V_m$
		\end{algorithmic}	
\end{algorithm}
\vspace{-5pt}
\section{Simulation Results}
\label{Results}
For simulations, we consider a rural small cell network in which the SBS is located at the center and $N_u$ users are deployed randomly within a circle of radius $800$~m. The separation between the SBS and core\footnote{If the separation between SBS and core is huge, relay nodes or macroBS may be deployed to relay the backhaul data to the core.} is set to $3$~km. The battery capacity $E_{\text{max}}$ is $2$~J and the maximum transmit power of the SBS, $P_{\text{max}}$ is $0.8$~W. The amount of harvested energy per arrival is set to $q=80$~mJ, while the time slot duration is set to $T=1$~ms. The total number of files that the users can request, $N$, is assumed to be very large for practical purposes. These files are assumed to be of same size and the cache size $M$ represents number of files cached. The exponent $s$ in \eqref{Zipf} is set to $2$. The fading parameter $\mu$ in \eqref{expDist} is $1$. The constants $a$ and $b$ in \eqref{funcg} are set to $0.18$ and $0.03$ respectively. The discount factor $\delta = 0.7$. The minimum SNR for backhaul, $\gamma_{\text{min}}$ is set to $8$~dB. The value of path loss exponent,$\alpha$, is set to $3$. The noise variance ${\sigma}^2$ is set to $-90$~dBmW. Since the focus of this work is to understand the impact of energy harvesting and caching on the downlink resource allocation at SBS, for simplicity, we assume equal bandwidth, $W=100$~kHz, for each user and backhaul.

\begin{figure}[!t]%
\includegraphics[width=\columnwidth]{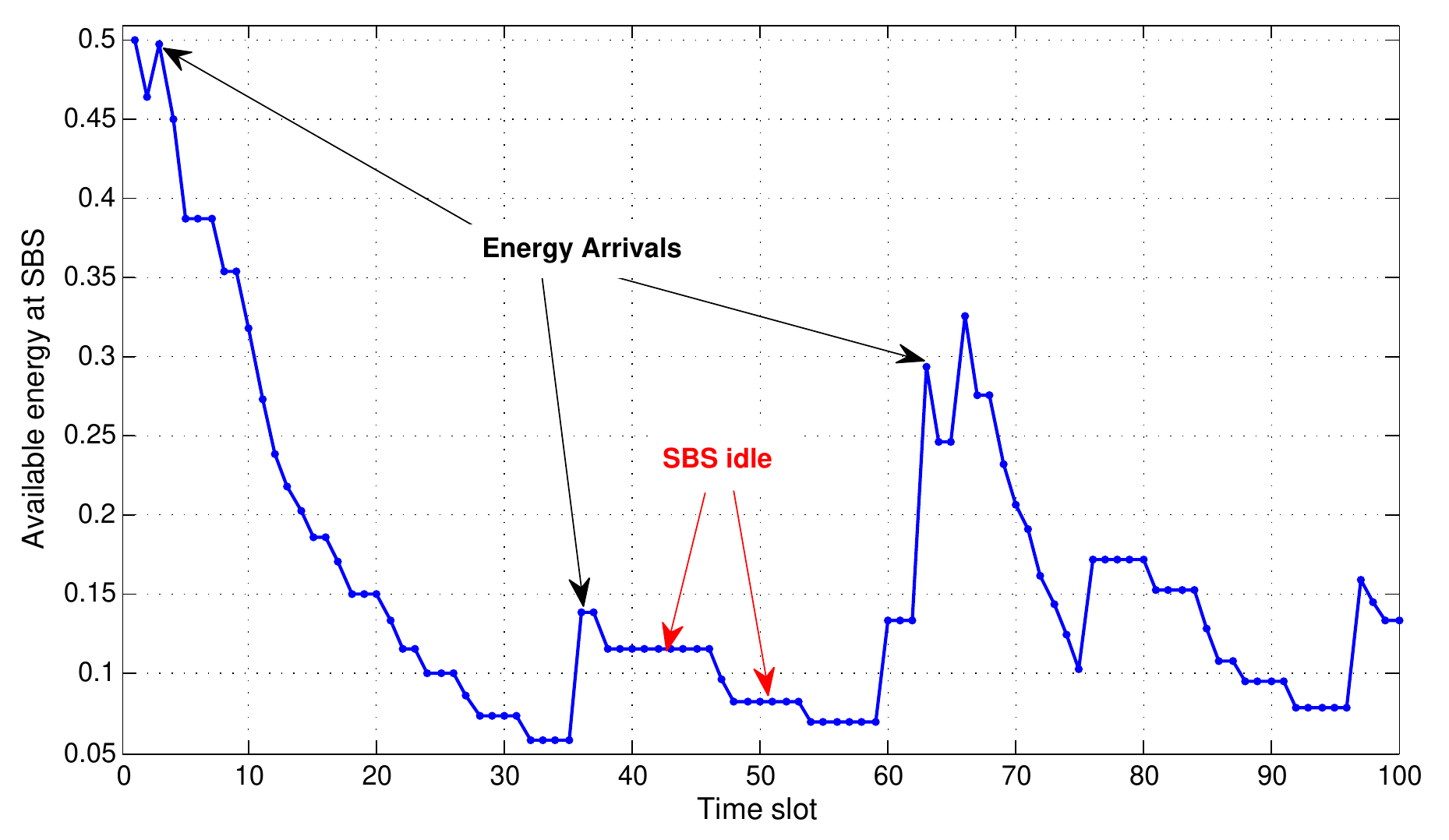}%
\vspace{-4pt}
\caption{Energy dynamics for $M=80$ and $\lambda = 0.1$.}%
\vspace{-13pt}
\label{energyDyn}%
\end{figure}

Figure~\ref{energyDyn} shows the variations in the energy available at SBS over time. The available energy decreases monotonously with time except for the time instants with energy arrivals or low battery level. When the battery level is low, the SBS enters energy-saving mode i.e., it remains idle when channel conditions are bad and until more energy is harvested.The rate of decrease in energy decreases with decrease in available energy at the SBS. This is due to the function $g(E_k)$ in  in \eqref{utility} that makes the SBS more energy-efficient when the battery level is low. Figure~\ref{energyDyn} also indicates the energy arrivals and time slots for which the SBS is idle (no transmission). We note that energy arrivals are infrequent because $\lambda$ is small. 

\begin{figure}[!t]%
\includegraphics[width=\columnwidth]{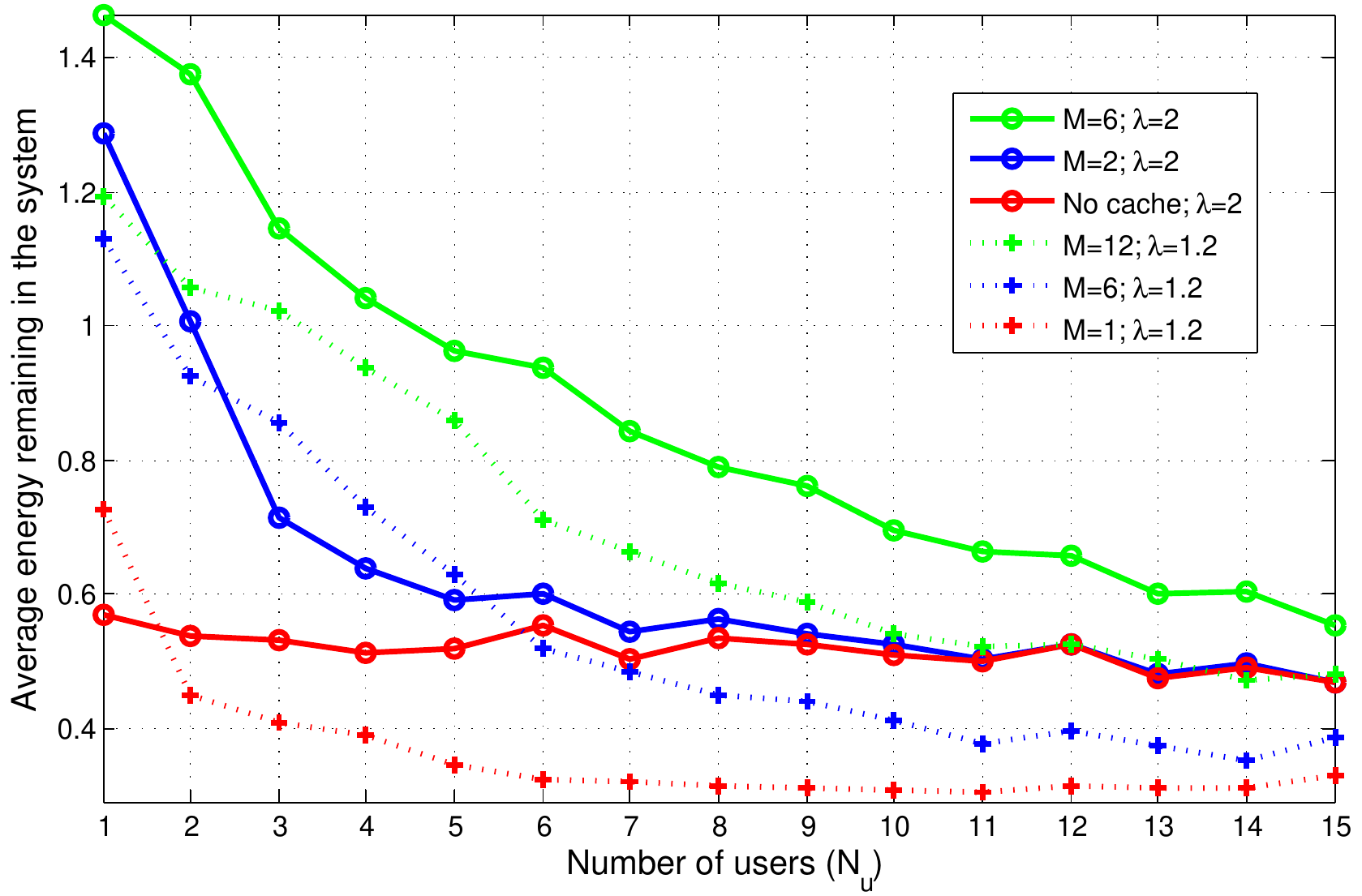}%
\caption{Impact of energy harvesting and cache size on available energy.}%
\vspace{-17pt}
\label{energySave}%
\end{figure}

Figure~\ref{energySave} shows the plot of average energy \emph{remaining} in the battery as a function of the number of users for different cache size $M$ and energy harvesting rate $\lambda$. We note that for a given energy harvesting rate, the available energy in the battery decreases sharply as the cache size is decreased. This is because the probability of miss, $\epsilon_m$ increases with the decrease in cache size. Therefore, the SBS will start using the backhaul more aggressively and, thus the energy expenditure increases. For $N_u=10$ and $\lambda=2~{\text{s}}^{-1}$, caching two files as opposed to none, results in $40\%$ higher energy savings. In Figure~\ref{energySave}, we also note that, as $N_u$ increases, the energy expenditure increases due to higher probability of backhaul usage, given by $\epsilon=1-{(1-\epsilon_m)}^{N_u}$. Finally, we note that the degradation in (energy) performance due to the decrease in energy arrival rate ($\lambda:2\rightarrow1.2$) can be compensated by increasing the cache size ($M:2\rightarrow6\rightarrow12$). 

\begin{figure}[!t]%
\includegraphics[height=0.7\columnwidth,width=\columnwidth]{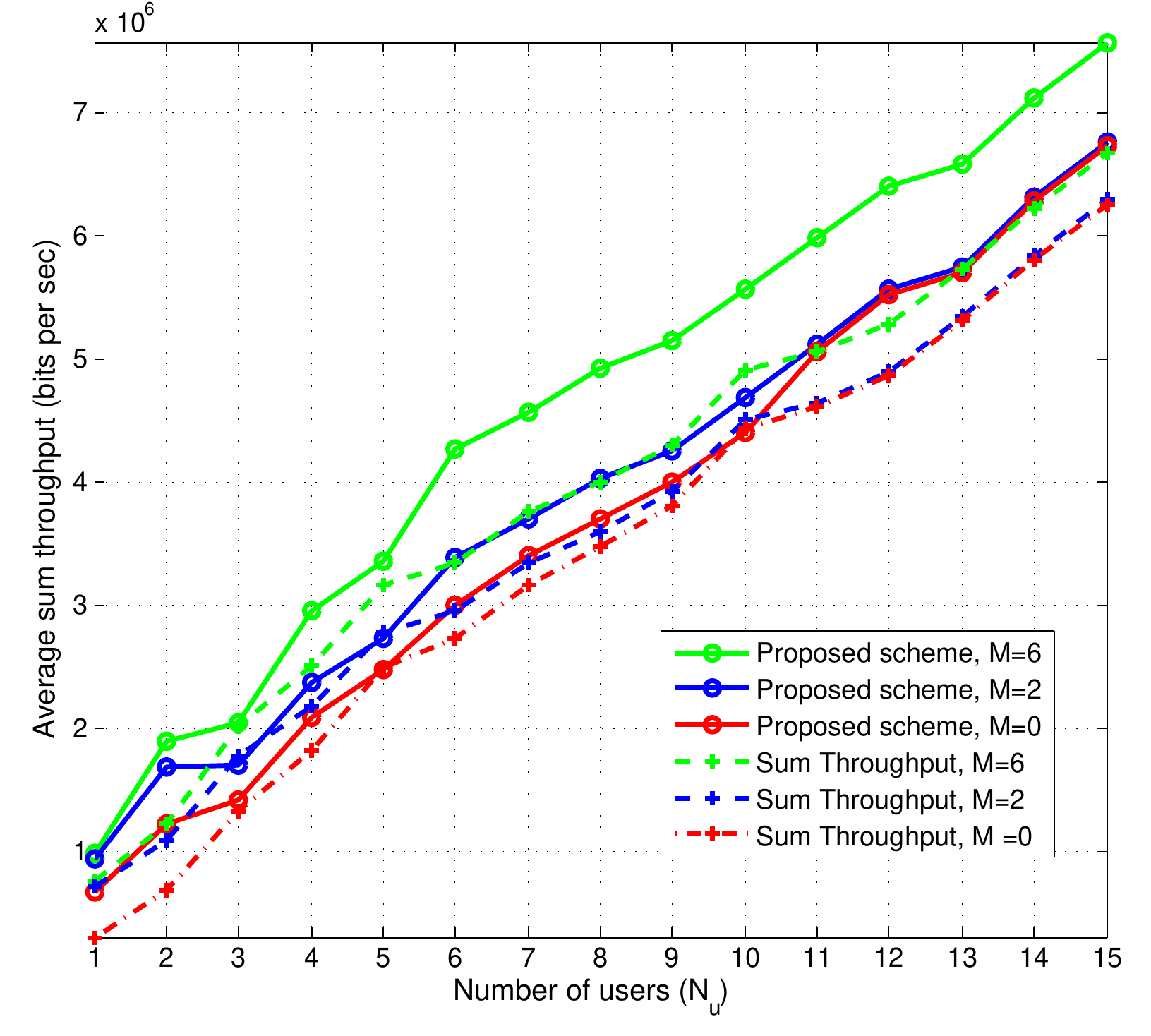}%
\caption{Impact of energy harvesting and caching on throughput.}%
\vspace{-5pt}
\label{DLThruput}%
\end{figure}

Figure~\ref{DLThruput} shows the average sum throughput of users as the number of users varies for different cache size. The rate of energy arrivals is $\lambda=2~{\text{s}}^{-1}$. We note that sum throughput increases with the increase in cache size. This is due to the fact that the frequency of backhaul access becomes smaller and thus more power is available for downlink transmissions. In this figure, we also compare the throughput performance of our proposed power control scheme with a sum throughput maximization scheme\footnote{The sum throughput maximization scheme can be obtained from the utility function defined in \eqref{utility} by setting the denominator to $1$.}. We note that, contrary to intuition, our scheme results in higher throughput ($20\%$ gain at $(N_u,M)=(11,6)$) than the sum throughput maximization scheme. This is because in case of sum throughput maximization scheme, the SBS transmits at its maximum power level while being oblivious to the channel state information. In the case of a deep fade, this will exhaust the battery power. If this is followed by a period of low energy arrivals, then the SBS will not be able to transmit at high power levels even in excellent channel conditions. This results in a significant loss in the sum throughput compared to the proposed scheme.  

\begin{figure}[!t]%
\includegraphics[width=\columnwidth]{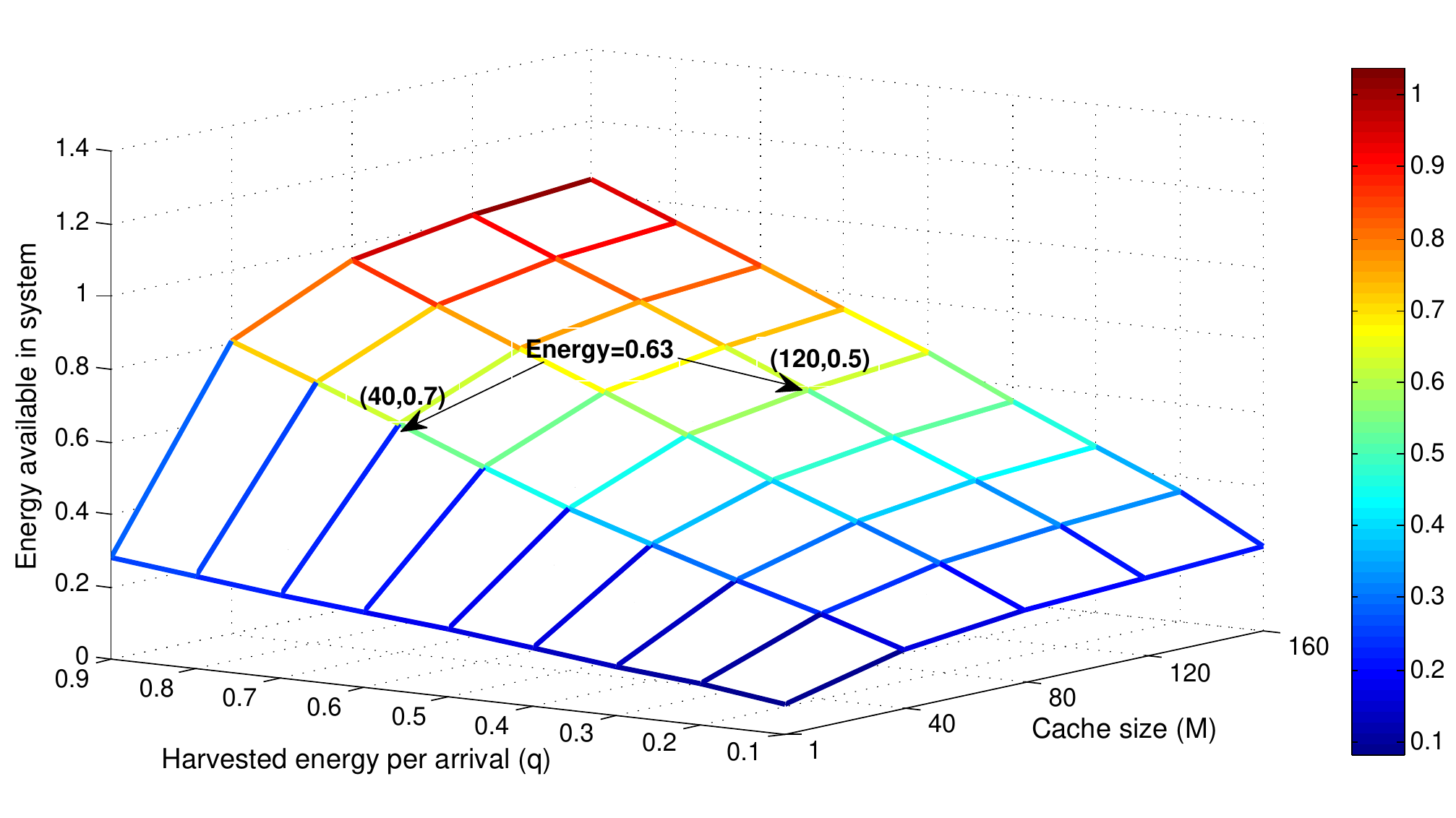}%
\caption{Effect of cache size and energy harvesting on energy available in system.}%
\vspace{-14pt}
\label{meshPlot}%
\end{figure}

In Figure~\ref{meshPlot}, we show the impact of jointly varying the cache size $M$ and the energy harvested per arrival $q$ on the energy availability at SBS. $N_u$ is set to $15$. We note that increasing $M$ from $1$ to $160$ or $q$ from $0.1$ to $0.9$, increases the energy availability. Increasing $q$ however has a more pronounced effect than increasing $M$. This is because we cache files in order of decreasing popularity, so increasing $M$ results in diminishing returns. In order to achieve a desired (energy) performance, the cache size and energy harvesting capacity can be traded off. This is illustrated in Fig.~\ref{meshPlot} by considering the points $(M,q) = (40,0.7)$ and $(120,0.5)$ which correspond to roughly same available energy of $0.63$~J. The practical implication of this result is that a network operator can make a better decision (after factoring in the storage and energy harvesting equipment costs) about the desired cache size and size of energy harvesting device that achieves desired performance.
\vspace{-5pt}
\section{Conclusions}
\label{concl}
In this paper, we have presented an energy efficient power control scheme for an energy harvesting SBS equipped with a wireless backhaul and local storage. We have formulated the power control problem as a discounted infinite horizon dynamic programming problem and solved it numerically using value iteration algorithm. Using simulations, we have revealed key insights on the impact of caching and energy harvesting on the energy consumption at SBS as the number of users in system is varied. Increasing cache size or energy harvesting rate results in increased energy availability. The proposed scheme also provides higher throughput as compared to the baseline sum throughput maximization scheme. We have also shown that the cache size can be bargained for energy harvesting capability and vice versa, while still meeting the desired system performance.

\bibliographystyle{IEEETran}	

\bibliography{ICCWorkshop}	

\begin{thebibliography}{10}
\providecommand{\url}[1]{#1}
\csname url@samestyle\endcsname
\providecommand{\newblock}{\relax}
\providecommand{\bibinfo}[2]{#2}
\providecommand{\BIBentrySTDinterwordspacing}{\spaceskip=0pt\relax}
\providecommand{\BIBentryALTinterwordstretchfactor}{4}
\providecommand{\BIBentryALTinterwordspacing}{\spaceskip=\fontdimen2\font plus
\BIBentryALTinterwordstretchfactor\fontdimen3\font minus
  \fontdimen4\font\relax}
\providecommand{\BIBforeignlanguage}[2]{{%
\expandafter\ifx\csname l@#1\endcsname\relax
\typeout{** WARNING: IEEEtran.bst: No hyphenation pattern has been}%
\typeout{** loaded for the language `#1'. Using the pattern for}%
\typeout{** the default language instead.}%
\else
\language=\csname l@#1\endcsname
\fi
#2}}
\providecommand{\BIBdecl}{\relax}
\BIBdecl

\bibitem{NSN11}
N.~S. Networks, ``2020: Beyond 4g – radio evolution for the gigabit
  experience,'' in \emph{White Paper}, 2011.

\bibitem{Damn11}
A.~Damnjanovic, J.~Montojo, Y.~Wei, T.~Ji, T.~Luo, and M.~Vajapeyam, ``A survey
  on 3gpp heterogeneous networks,'' \emph{IEEE Wireless Communications},
  vol.~18, no.~3, pp. 10--21, 2011.

\bibitem{smallCellForum}
\BIBentryALTinterwordspacing
``Extending rural and remote coverage using small cells,'' Small Cell Forum,
  Tech. Rep. 047.04.01, Feb. 2014. [Online]. Available:
  \url{http://scf.io/en/documents/047_Extending_rural_and_remote_coverage_using_small_cells.php}
\BIBentrySTDinterwordspacing

\bibitem{Matthias13}
M.~Wildemeersch, T.~Quek, C.~Slump, and A.~Rabbachin, ``Cognitive small cell
  networks: Energy efficiency and trade-offs,'' \emph{IEEE Transactions on
  Communications}, vol.~61, no.~9, pp. 4016--4029, Sep. 2013.

\bibitem{Sabella14}
D.~Sabella, M.~Caretti, and R.~Fantini, ``Energy saving schemes for
  self-backhauled small cells in lte-advanced networks,'' in \emph{IEEE
  Wireless Communications and Networking Conference Workshops}, Apr. 2014, pp.
  23--28.

\bibitem{zhangLau14}
F.~Zhang and V.~Lau, ``Closed-form delay-optimal power control for energy
  harvesting wireless system with finite energy storage,'' \emph{IEEE
  Transactions on Signal Processing}, vol.~62, no.~21, pp. 5706--5715, Nov
  2014.

\bibitem{huang14}
\BIBentryALTinterwordspacing
X.~Huang and N.~Ansari, ``Optimal cooperative power allocation for energy
  harvesting enabled relay networks,'' \emph{arXiv}, vol. abs/1405.5764, 2014.
  [Online]. Available: \url{http://arxiv.org/abs/1405.5764}
\BIBentrySTDinterwordspacing

\bibitem{KwanLo13}
D.~Ng, E.~Lo, and R.~Schober, ``Energy-efficient resource allocation in ofdma
  systems with hybrid energy harvesting base station,'' \emph{IEEE Transactions
  on Wireless Communications}, vol.~12, no.~7, pp. 3412--3427, July 2013.

\bibitem{gong14}
J.~Gong, J.~Thompson, S.~Zhou, and Z.~Niu, ``Base station sleeping and resource
  allocation in renewable energy powered cellular networks,'' \emph{IEEE
  Transactions on Communications}, vol.~62, no.~11, pp. 3801--3813, Nov 2014.

\bibitem{molisch13}
N.~Golrezaei, A.~Molisch, A.~Dimakis, and G.~Caire, ``Femtocaching and
  device-to-device collaboration: A new architecture for wireless video
  distribution,'' \emph{IEEE Communications Magazine}, vol.~51, no.~4, Apr.
  2013.

\bibitem{Golrezaei12}
N.~Golrezaei, K.~Shanmugam, A.~Dimakis, A.~Molisch, and G.~Caire,
  ``Femtocaching: Wireless video content delivery through distributed caching
  helpers,'' in \emph{IEEE INFOCOM}, March 2012, pp. 1107--1115.

\bibitem{Tadrous13}
J.~Tadrous, A.~Eryilmaz, and H.~El~Gamal, ``Proactive content download and user
  demand shaping for data networks,'' \emph{IEEE/ACM Transactions on
  Networking}, vol.~PP, no.~99, pp. 1--1, 2014.

\bibitem{Caire13}
M.~Ji, G.~Caire, and A.~Molisch, ``Fundamental limits of distributed caching in
  d2d wireless networks,'' in \emph{IEEE Information Theory Workshop}, Sept
  2013, pp. 1--5.

\bibitem{pingyod14}
A.~Pingyod and Y.~Somchit, ``Content updating method in femtocaching,'' in
  \emph{International Joint Conference on Computer Science and Software
  Engineering (JCSSE)}, May 2014, pp. 123--127.

\bibitem{Ozel11}
\BIBentryALTinterwordspacing
O.~Ozel, K.~Tutuncuoglu, J.~Yang, S.~Ulukus, and A.~Yener, ``Transmission with
  energy harvesting nodes in fading wireless channels: Optimal policies,''
  \emph{arXiv}, vol. abs/1106.1595, 2011. [Online]. Available:
  \url{http://arxiv.org/abs/1106.1595}
\BIBentrySTDinterwordspacing

\bibitem{Dhillon14}
H.~Dhillon, Y.~Li, P.~Nuggehalli, Z.~Pi, and J.~Andrews, ``Fundamentals of
  heterogeneous cellular networks with energy harvesting,'' \emph{IEEE
  Transactions on Wireless Communications}, vol.~13, no.~5, pp. 2782--2797, May
  2014.

\bibitem{Gelenbe14}
E.~Gelenbe, ``A sensor node with energy harvesting,'' \emph{SIGMETRICS},
  vol.~42, no.~2, pp. 37--39, Sep 2014.

\end{thebibliography}

\end{document}